\documentclass[prx,
superscriptaddress,
twocolumn,
amsmath,amssymb
]{revtex4-2}

\usepackage{placeins}
\usepackage{graphicx}% Include figure files
\usepackage{dcolumn}% Align table columns on decimal point
\usepackage{bm}% bold math
\usepackage{lipsum}
\usepackage{braket}
\usepackage[english]{babel}
\usepackage{amsmath}
\usepackage{xcolor}

\DeclareOldFontCommand{\bf}{\normalfont\bfseries}{\mathbf}
\DeclareOldFontCommand{\rm}{\normalfont\rmseries}{\mathrm}
\DeclareOldFontCommand{\tt}{\normalfont\ttseries}{\mathtt}

\DeclareMathAlphabet{\mathup}{OT1}{\familydefault}{m}{n} % only for non-lualatex
\newcommand*{\mup}[1]{\mathup{#1}} % Abkürzung für \mathup

\newcommand*{\iu}{\mathup{i}\mkern1mu}

\DeclareMathOperator{\e}{e}

\newcommand\ten[1]{\mathbf{#1}}
\newcommand\dirvec[1]{\mathbf{e}_{#1}}

\begin{document}

\title{Viscoelastic dynamics of a soft strip subject to a large deformation}

\author{Alexandre Delory}
\email[Corresponding author: ]{alexandre.delory@espci.psl.eu}
\affiliation{ Institut Langevin, ESPCI Paris, Université PSL, CNRS, 75005 Paris, France }%
\affiliation{ Physique et M\'ecanique des Milieux H\'et\'erog\`enes, CNRS, ESPCI Paris, Universit\'e PSL, Sorbonne Universit\'e, Université de Paris Cit\'e, F-75005}%
\author{Daniel A. Kiefer}
\affiliation{ Institut Langevin, ESPCI Paris, Université PSL, CNRS, 75005 Paris, France }%
\author{Maxime Lanoy}
\affiliation{ Laboratoire d'Acoustique de l'Université du Mans (LAUM), UMR 6613, Institut d'Acoustique - Graduate School (IA-GS), CNRS, Le Mans Université, 72085 Le Mans, France.}%
\author{Antonin Eddi}
\affiliation{ Physique et M\'ecanique des Milieux H\'et\'erog\`enes, CNRS, ESPCI Paris, Universit\'e PSL, Sorbonne Universit\'e, Université de Paris Cit\'e, F-75005}%
\author{Claire Prada}
\affiliation{ Institut Langevin, ESPCI Paris, Université PSL, CNRS, 75005 Paris, France }%
\author{Fabrice Lemoult}
\affiliation{ Institut Langevin, ESPCI Paris, Université PSL, CNRS, 75005 Paris, France }

%\date{\today}

\begin{abstract}
To produce sounds, we adjust the tension of our vocal folds to shape their properties and control the pitch. This efficient mechanism offers inspiration for designing reconfigurable materials and adaptable soft robots.
However, understanding how flexible structures respond to a significant static strain is not straightforward. This complexity also limits the precision of medical imaging when applied to tensioned organs like muscles, tendons, ligaments and blood vessels among others.
In this article, we experimentally and theoretically explore the dynamics of a soft strip subject to a substantial static extension, up to 180\%. Our observations reveal a few intriguing effects, such as the resilience of certain vibrational modes to a static deformation. These observations are supported by a model based on the incremental displacement theory. 
This has promising practical implications for characterizing soft materials but also for scenarios where external actions can be used to tune properties.
\end{abstract}

\maketitle

%%% MAIN TEXT %%%%
\section*{Introduction}

Soft solids exhibit the remarkable ability to undergo substantial deformations.
This unique property is harnessed by living organisms, notably allowing them to achieve locomotion~\cite{josephson1993contraction,huffard2005underwater,wilkinson2016restless} but also enabling the whole morphogenetic chain to happen, through the ability of shaping tissues, and leading to the development of organs and physiological functionalities~\cite{dervaux2008morphogenesis,heisenberg2013forces,goriely2017mathematics}. Similarly, plants leverage this trait to adapt to varying environmental conditions~\cite{liang_2011,goriely2017mathematics,moulia2021fluctuations}.
With recent scientific advances in the fields of gels and polymers~\cite{mondal2020review,kaspar2021rise,zhao2021soft}, engineers have achieved significant progress in the manufacturing of versatile, biocompatible and durable soft structures. This has led to the development of soft robots~\cite{laschi2016soft,mustaza_2019,majidi2019soft,ng2021locomotion}, medical devices~\cite{chin2017additive,li2022miniature,smith2023soft}, and inflatable structures~\cite{siefert2020programming,gao2023pneumatic,tanaka2023turing}, among others. 

The unique mechanical behavior of soft solids arises from their microstructure. Biological tissues consist of large biomolecules, primarily proteins but also nucleic acids, linked together by covalent bonds (such as peptide and phosphodiester bonds) and non-covalent interactions (including hydrogen bonds, van der Waals forces, and ionic interactions). These networks possess a random nature and offer residual configurational freedom at multiple scales. As a result, the elasticity is governed by entropy~\cite{treloar1975physics, bustamante1994entropic, boyce2000constitutive}, which usually results in a non-Hookean response. This effect can be rendered by considering a Young's modulus $E$, which becomes dependent on the deformation; a feature of hyperelastic materials. Furthermore, spatial rearrangements induce dissipation through viscous effects, leading to relaxation phenomena. This suggests that the Young's modulus is both complex-valued and frequency-dependent.

These characteristics can be illustrated using a model system. Here, we consider a simple strip. It is made of a commercially available silicone rubber, the Smooth On Ecoflex 00-30. Such elastomers consist of an entanglement of macromolecules connected together through the action of a cross-linking agent. Therefore, it is relevant to draw an analogy between their phenomenology and that of biological tissues.

\begin{figure}
    \centering
    \includegraphics[width=.49\textwidth]{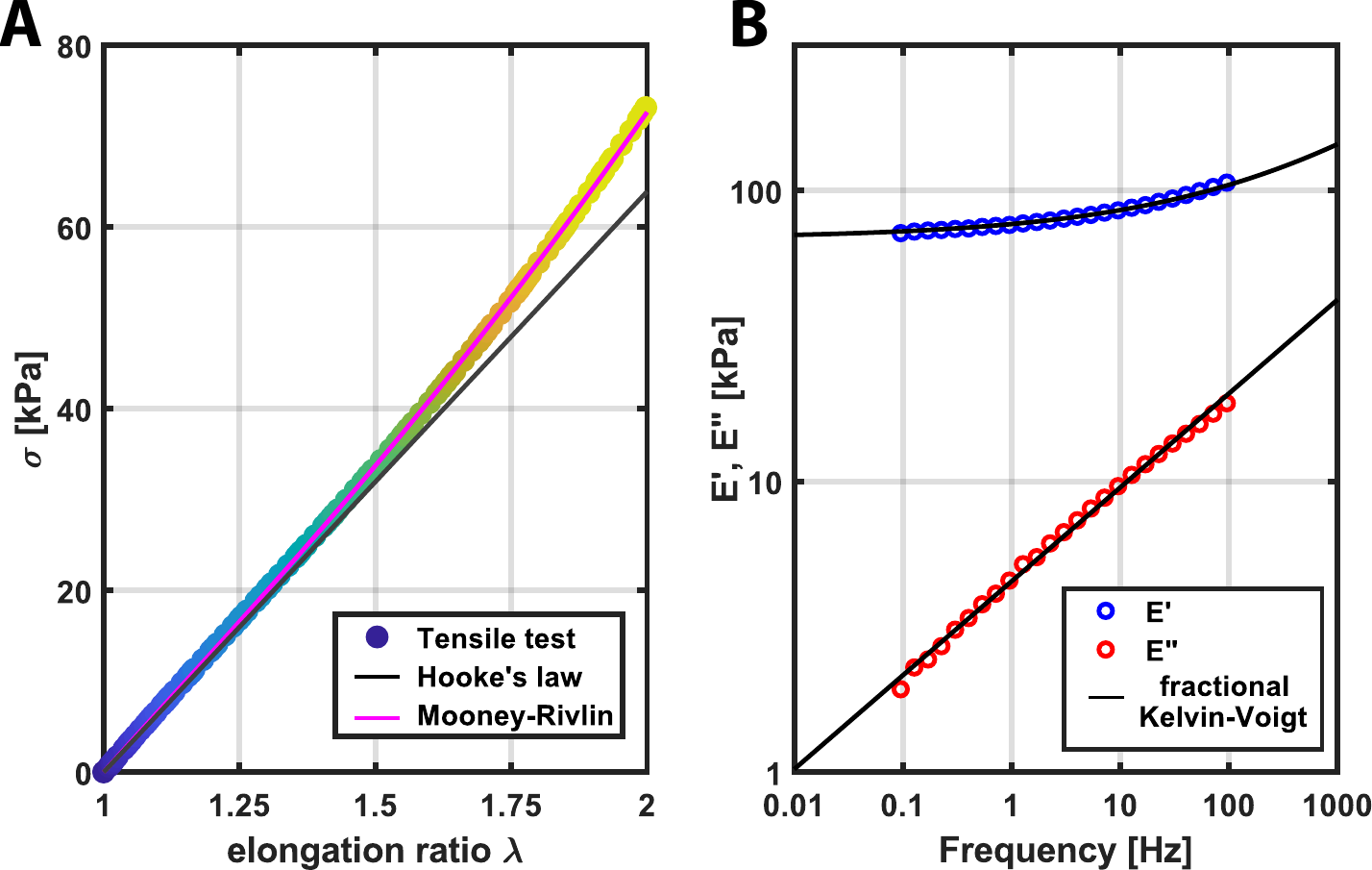}
    \caption{\textbf{Mechanical characterization of Ecoflex 00-30 ---} (A) Experimental tensile test (symbols) and predictions (solid lines) using a linear (Hooke law, in black) and a nonlinear model (Mooney-Rivlin, in magenta). The elongation ratio $\lambda$ is measured as the ratio of the deformed total length with respect to the initial total length. (B) Experimental characterization (symbols) of the complex Young's modulus together with its fit (solid lines) based on a fractional Kelvin-Voigt model.}
    \label{fig:Rheology}
\end{figure}

Just like biological tissues, silicone elastomers are expected to exhibit a hyperelastic behavior~\cite{treloar_2005,arrudaboyce_1993}. To verify it, we conducted a tensile test on an Ecoflex strip. Results, gathered in Figure~\ref{fig:Rheology}A, indicate that for a stretch ratio $\lambda$ (ratio of deformed to original length) larger than 1.5, the material response deviates from Hooke's law. This deviation can be captured using the 2-parameter incompressible Mooney-Rivlin model~\cite{mooney1940theory,rivlin1948large,beatty2001seven,puglisi_2016} which links the Cauchy stress $\sigma$ to the stretch ratio $\lambda$ as:

\begin{equation}\label{eq:MooneyRivlin}
    \sigma = \frac{E_0}{3} \left(1-\alpha+\frac{\alpha}{\lambda}\right) \left(\lambda^2-\frac{1}{\lambda}\right).
\end{equation}
with the Mooney-Rivlin parameters $E_0$ and $\alpha$. This equation recovers Hooke's law $\sigma=E_0\left(\lambda-1\right)$ in the limit of small elongations.

This formula effectively depicts the elastic response of an Ecoflex strip within the investigated elongation range for $E_0=67$~kPa and $\alpha=0.15$. By assuming Hooke's law and measuring the slope of the curve at small elongations $\lambda\le1.3$, the value of $E_0=67$~kPa is again recovered.
From the Mooney-Rivlin model, one can derive an effective Young's modulus (see details in Appendix A), which amounts to considering the material as Hookean with the following elongation-dependent Young's modulus:

\begin{equation}\label{eq:Young_MooneyRivlin}
    E(\lambda,\omega=0) = \frac{E_0}{3} \left[ (1-\alpha)\left(1+\frac{2}{\lambda^3}\right) + \frac{3\alpha}{\lambda^4}\right].
\end{equation}
\noindent Nevertheless, this expression is only valid in the static regime at a zero angular frequency $(\omega = 0)$.

Another consequence of the polymer chains' ability to conform freely is that they rearrange with a characteristic relaxation time. As a result, the Young's modulus is expected to be frequency-dependent and complex-valued ($E=E'+\text{i} E''$).
By performing oscillatory rheological measurements using a plate-plate geometry \footnote{This apparatus actually probes the shear modulus $G$ from which the Young's modulus is extracted as $E=3G$ for incompressible solids (Poisson's ratio $\nu\approx1/2$).}, we also investigated the frequency dependence of our silicone rubber. 
As shown in Figure~\ref{fig:Rheology}B, the material response appears to be essentially elastic in the range of frequencies tested with our apparatus ($E'\gg E''$). The loss modulus $E''$ grows as a power law of frequency $\omega$, which is a signature of a visco-elastic material. These trends can be modeled using a fractional derivative Kelvin-Voigt law~\cite{smit1970rheological,bagley1983fractional,liu_2014,rolley2019flexible,mainardi2022fractional,sharma_2023} (solid lines) which writes:

\begin{equation}\label{eq:rheology}
   E(\lambda=1,\omega)=E_0\left[1+(i\omega\tau)^n\right].
\end{equation}

This model convincingly fits our data (see Figure~\ref{fig:Rheology}B) with the following parameters: $E_0=69$~kPa, $\tau=330$~$\mu$s, and $n=0.32$. Note that the value of $E_0$ obtained here matches that of the tensile test (Figure~\ref{fig:Rheology}A); slight difference comes from the utilization of distinct samples for the two measurements. We emphasize that the above expression is valid in the absence of external deformation $(\lambda=1)$.

In summary, the dynamic response of our soft polymer depends on both frequency and strain, which are usually examined separately. However, drawing conclusions about $E$ for any couple of parameters $(\lambda, \omega)$ is not straightforward due to their interdependence. Until recently~\cite{delory_eml_2023}, there has been no unified framework to account for both simultaneously.
As a consequence, capturing the dynamics of soft structures under significant stress, a common occurrence in our daily environment (vocal folds, tendons, ligaments, muscles and blood vessels are a few examples of organs operating dynamically under stress), remains a challenge.

In this article, we propose a model system to explore the interplay between static prestress and dynamic response of a soft structure. Specifically, we investigate the behavior of a rubber strip under significant static deformation. Using image correlation techniques, we track the propagation of a small perturbation and make several intriguing discoveries. For instance, we find that compressional waves exhibit remarkable resilience to external stretching, while flexural modes display heightened sensitivity.
To support our findings, we derive a comprehensive model adapted from the principles of acousto-elasticity~\cite{ogden_1993,rogerson_1995,ogden_1997,nolde_2004,saccomandi_2003,destrade_2007,colonnelli_2013,berjamin_2022,patra_2023}. Finally, we examine the effect of stretching on Dirac cones, drawing an analogy with condensed matter physics~\cite{ashcroft2022solid}.
Here, we find that the cone is relatively immune to longitudinal stretching, but breaks appart upon transversal stretching.
Overall, our work illustrates how external stress can be harnessed to tune the dynamics of soft materials, offering applications for the design of adaptive structures~\cite{khajehtourian2021soft} or tunable metamaterials~\cite{barnwell_2016,miniaci_2021,berjamin_2022}.

\section*{Low-frequency dynamics}
\subsection*{Experimental set-up}

\begin{figure*}
    \centering
    \includegraphics[width=\textwidth]{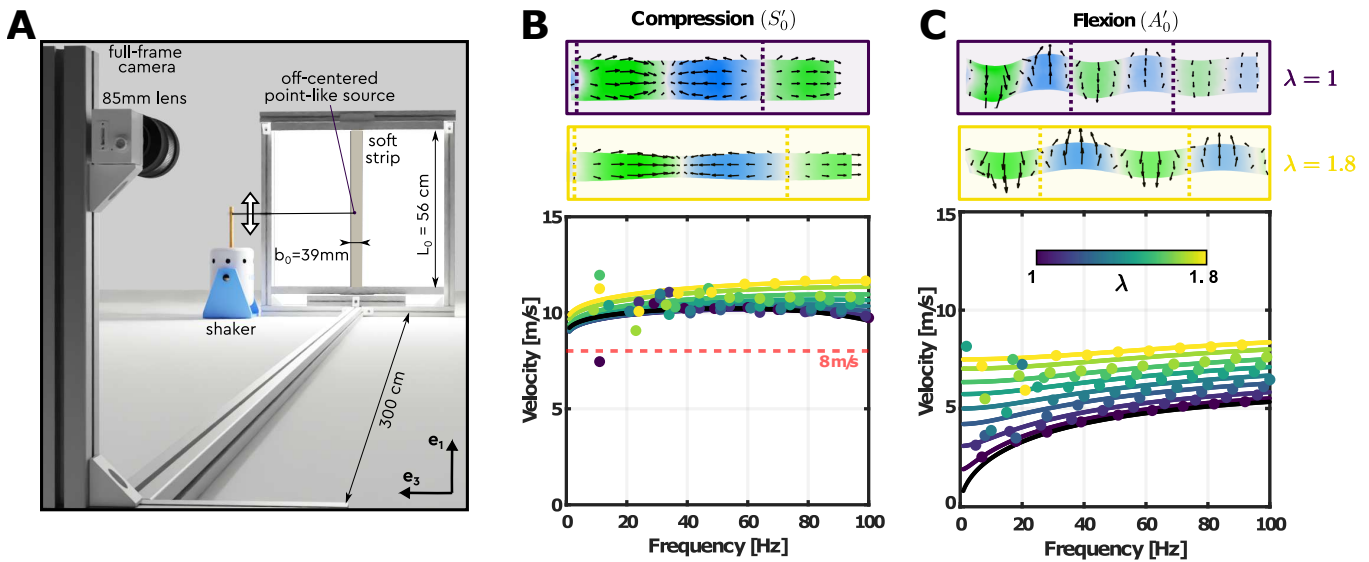}
    \caption{\textbf{Tracking in-plane waves in a stretched strip --- } (A) Experimental set-up: a strip of dimensions $L_0\times b_0 \times h_0$ made of a soft elastomer (Ecoflex 00-30) is stretched by a factor $\lambda$ in the $\ten{e_1}$ direction. Elastic waves are generated by a shaker and the strip motion is tracked by a remote camera. By adequate post-processing, the motion is decomposed into the two vibrational eigenmodes, \textit{i.e.} compression and flexion. (B) Top: profiles of the displacement at 50~Hz corresponding to a compressional mode for $\lambda=1$ (blue frame) and $\lambda=1.8$ (yellow frame). The dashed lines indicate the wavelength. Bottom: Phase velocity of the compressional wave for static elongations varying from $\lambda=1$ (blue) to $\lambda=1.8$ (yellow). Experimental measures (points) are compared with theoretical predictions (lines) using our three-dimensional model. (C) Same as B but for the flexural mode.}
    \label{fig:LowFrequency}
\end{figure*}

To monitor the motion in a strip, we designed the experimental set-up sketched on Figure~\ref{fig:LowFrequency}A. 
A strip of initial length $L_0=56$~cm, width $b_0=39$~mm and thickness $h_0= 3$~mm, made of a soft elastomer (Ecoflex 00-30) is hang up and then stretched by a factor $\lambda$ in the longitudinal direction~$\ten{e_1}$.
Wave generation is performed by a shaker (TIRAVib 51120) driven in the harmonic regime. The source is set so that the motion remains in the ($x_1,x_3$) plane. In this configuration, the displacement field is tracked by a standard video camera facing the strip. For a given angular frequency $\omega$, a Digital Image Correlation (DIC) algorithm and a time-Fourier coefficient derivation are implemented to extract the complex-valued in-plane components of the field $u_1(\omega,x_1,x_3)$ and $u_3(\omega,x_1,x_3)$.
Details and typical movies can be found in Ref.~\cite{lanoy_PNAS_2020,delory_JASA_2022}. A singular value decomposition is then employed to separate contributions from the fundamental vibrational eigenmodes. 
Assuming that the motion is purely polarized in the ($x_1,x_3$) plane, only two modes contribute at low frequency: a compressional mode mostly polarized in the $\ten{e_1}$ direction and a flexural mode mostly polarized in the $\ten{e_3}$ direction.
Displacement profiles obtained at 50~Hz are displayed in Figure~\ref{fig:LowFrequency}B and C. They show that the wavelength (see dashed lines) increases as the stretch ratio $\lambda$ increases from $1$ (blue frame) to $1.8$ (yellow frame). This effect is more pronounced in the case of the flexural mode.

From these measurements, one can estimate the respective phase velocities, \textit{i.e.} the product of wavelength and frequency. Experimental results are reported as symbols in Figure~\ref{fig:LowFrequency}B and C. Here, the applied stretch varied from 1 to 1.8 and the driving frequency from 0 to 100~Hz. The experimental data in Figure~\ref{fig:LowFrequency} are compared with numerical ones, based on a procedure that is described along with our three-dimensional model.
These measurements indicate that the flexural mode is highly affected by the application of a static longitudinal tension, while the compressional mode appears nearly immune to it. As a matter of fact, the latter travels at roughly 10~m/s, regardless of both $\lambda$ and $\omega$. To clarify these observations, let us examine analytical expressions for these two velocities.

\subsection*{Compression} 
In the fields of structural mechanics and biomedical imaging, it is common to employ the static Young's modulus $E_0$, to make inferences about the dynamic behavior of a given structure. For soft materials, this approach leads to substantial errors. To illustrate this point, one can go back to the equation governing compressional dynamics (see for instance equation (25.1) in reference~\cite{landau1986theory}):

\begin{equation}\label{eq:LongiWave}
    E\frac{\partial^2u_1}{\partial {x_1}^2}-\rho \frac{\partial^2u_1}{\partial t^2}=0,
\end{equation}

\noindent with $\rho$ standing for the material mass density.
Estimating the phase velocity from the static Young's modulus $E_0$ and a mass density $\rho=1000$ kg/m$^3$ leads to a value of $V_c^0=\sqrt{E_0/\rho}=8~$m/s (indicated by a red dashed line in Figure~\ref{fig:LowFrequency}B) which significantly underestimates observations.
In addition, this approach assumes a constant velocity, \textit{i.e.} non-dispersive propagation, which is not entirely accurate in this case.

A better estimation can be obtained by considering the material's rheology and replacing the Young's modulus with that of equation~(\ref{eq:rheology}). In the frequency range investigated here, since one always has $\omega\tau \ll 1$, the phase velocity writes:

\begin{equation}
\label{eq:VCompressionRheology}
    V_c(\lambda=1,\omega)= V_c^0\left[1+(\omega\tau)^n \cos\left(n\pi /2 \right)/2\right].
\end{equation}

Because we take into account the rheology, this amounts to a frequency dependent phase velocity as experimentally observed. At 50~Hz, this expression yields a velocity of roughly 10~m/s, in much better agreement with our measurements. Also, this velocity slowly grows in power law with frequency.

Our experiments also indicate that $V_c$ is almost independent of the applied stretch $\lambda$. To capture this effect accurately, it is necessary to incorporate the hyperelastic prediction for the Young's modulus from equation~(\ref{eq:Young_MooneyRivlin}). However, this alone is not enough.
In fact, wave equation~(\ref{eq:LongiWave}) is written with undeformed coordinates and a push-forward operation is required to obtain the correct Young's modulus. This operation amounts to transitioning from a Lagrangian (material coordinates) to an Eulerian (laboratory coordinates) description.
In the present simple uniaxial configuration, it corresponds~\cite{zhao2021elastic} to replacing the Young's modulus by $\lambda^2 E$. 
Overall, the compressional velocity can be found by replacing the expression of $V_c^0$ in equation~(\ref{eq:VCompressionRheology}) by:

\begin{equation} \label{eq:Vcompression}
V_c^0(\lambda)=\sqrt{\frac{\lambda^2 E(\lambda,\omega=0)}{\rho}}.
\end{equation}

For values of $\lambda$ ranging from 1 to 1.8 and for the Mooney-Rivlin model from equation~(\ref{eq:Young_MooneyRivlin}), this predicts that $V_c^0$ undergoes variations of -2\% to +7\% maximum. 
Hence, this expression effectively captures the small variations observed in the experimental points of Figure~\ref{fig:LowFrequency}B, using simple physical arguments. 

Lastly, at $\lambda=1$ the velocity starts decreasing for frequencies higher than 50~Hz, which is not supported by this approach. At this frequency, the wavelength becomes comparable to the strip's width, and the one-dimensional model inevitably fails.

\subsection*{Flexion}

Unlike compression, the flexural dynamics displays a remarkable sensitivity to the application of a static stress. At 10~Hz, our measurements indicate that the velocity goes from 2~m/s at $\lambda=1$ up to 7~m/s at $\lambda=1.8$.
Interestingly, the static stress triggers a bifurcation in the dispersion behavior.
For instance, as illustrated in Figure~\ref{fig:LowFrequency}C, when $\lambda\approx 1$ (dark blue symbols), the flexural wave is highly dispersive. Its velocity grows from 0~m/s in the quasi-static limit to around 5~m/s at 100~Hz. In contrast, for $\lambda=1.8$, the velocity becomes nearly independent of frequency, \textit{i.e.} the propagation is non-dispersive.
This is characteristic of the transition from a flexural beam regime to a string-like regime, governed by the tension in the material. 

This effect can be captured by getting back to the simple Euler-Bernoulli model~\cite{doyle1989wave}:

\begin{equation}\label{eq:EulerBern}
    \frac{EI}{A}\frac{\partial^4u_3}{\partial {x_1}^4}-\sigma\frac{\partial^2u_3}{\partial {x_1}^2}+\rho \frac{\partial^2u_3}{\partial t^2}=0,
\end{equation}

\noindent with $A=bh$ the strip cross-sectional area, $I=hb^3/12$ the second moment of area~\footnote{Note that this moment corresponds to the bending in the $(x_1,x_3)$ plane.} and $\sigma$ the uniaxial stress due to the applied tension force. 
Assuming a propagative solution with wavenumber $k$, one obtains the following dispersion relation:

\begin{equation}
\frac{2EI}{A}k^2=-\sigma\pm\sqrt{\sigma^2+\frac{4EI}{A}\rho\omega^2}.
\end{equation} 

From this expression, it is possible to identify a non-dimensional parameter $\gamma=A\sigma^2/4\rho E I \omega^2$ which renders the competition between tension and bending, and evidences the existence of the two aforementioned regimes. When the strip is not stretched, $\sigma=0$ and $\gamma$ vanishes. The phase velocity of flexural waves writes:

\begin{equation}
V_{f}\left(\lambda=1,\omega\right)=\sqrt{\omega}\left[\frac{EI}{\rho A}\right]^{\frac{1}{4}}.
\label{eq:Vflexion}
\end{equation}

The $\sqrt{\omega}$ dependence is the signature of a strongly dispersive regime (depicted by the dark blue line in Figure~\ref{fig:LowFrequency}C). Interestingly, in this configuration, the group velocity is exactly twice the phase velocity; enabling possible analogies with non-relativistic free particle~\cite{broglie1925recherches}. 
Obviously, this velocity should saturate at some point; otherwise these waves would become infinitely fast. This highlights a limitation of the Euler-Bernoulli model, which becomes invalid at higher frequencies because it assumes that the displacement should remain purely transverse.
Besides, just like in the previous section, the question arises of which expression one should consider regarding the Young's modulus $E$. Very similarly, including rheology in equation~(\ref{eq:rheology}) yields a more accurate prediction for velocity.

As the stress $\sigma$ increases with $\lambda$ varying from 1.1 to 1.8, the non-dimensional parameter $\gamma$ grows from 0.1 to 8, and a change in the strip behavior is reached. 
Now, flexion is completely governed by the tension, and the velocity simply writes:

\begin{equation}
V_{f}\left(\lambda,\omega\right)=\sqrt{\sigma/\rho}.
\label{eq:flexVelo}
\end{equation}

This expression captures the almost non-dispersive dynamics observed in the low-frequency regime, but especially the increase in velocity with $\lambda$ in the limit $\omega \to 0$.
String instruments, like guitars or violins, precisely operate in this regime: the pitch produced by the musician is completely governed by the fine adjustment of the stress $\sigma$ in the strings. Some other instruments, like the xylophone or the glockenspiel, are designed for the first regime, where the pitch is typically controlled by the strip lengths rather than tension.

This uni-dimensional model provides an efficient picture of the strip behavior in the two asymptotic regimes. But for intermediate values of $\lambda$ the estimation of $V_f$ becomes more challenging. Besides, the models of flexion and compression do not take into account the finite size of the strip leading to inconsistencies when frequency increases, or more precisely, when the wavelength becomes comparable to the strip width. Finally, they do not offer the possibility to clearly evidence the respective roles of rheology and stretching, which happens to be crucial here.
All these considerations substantiate the need for constructing a comprehensive three-dimensional model.

%%%%%%%%%%%%%%%%%%%%%%% THEORY %%%%%%%%%%%%%%%%%%%%%%%%%%%%%%%%%%%%
\subsection*{Three-dimensional model} 
To build such a model, one has to come back to the constitutive equations of continuum mechanics~\cite{mindlin_1960,medick_1968,meleshko_2010,krushynska_2011}. Assuming a compressible and homogeneous material, the displacement field is solution of the equation of motion:

\begin{equation}
    C_{jikl} \frac{\partial^2 u_k}{\partial x_j\partial x_l} = \rho \frac{\partial^2 u_i}{\partial t^2},
    \label{eq:equation_of_motion_underformed}
\end{equation}

\noindent where the Einstein notation is used (sum over repeated indices $j,k,l$), $\ten{u}\left(\ten{x},t\right)$ is the local displacement vector with its 3 components $u_i$, and $C_{jikl}$ are the fourth-order elasticity tensor components~\cite{royer_1999,royer2022elastic}.

The effect of prestress is implemented thanks to the acoustoelastic theory~\cite{ogden_1997,destrade_2007}. It applies in the context of incremental displacements, \textit{i.e.} small dynamic perturbations on top of a large static deformation.
As explained in a recent contribution~\cite{delory_eml_2023}, this consists in replacing the stiffness tensor of the undeformed isotropic elastic material assuming the applied prestress is homogeneous and in the time-harmonic case, by a frequency and strain dependent elasticity tensor $\ten{C^\omega}$ whose components write:

%\small
\begin{equation}
    C^\omega_{jikl}(\lambda, \omega)\!=\! C^0_{jikl}(\lambda)\!+\!\frac{E_0}{3} {I}_{jikl}\!
    \left[ 1\!+\!\beta'\,\frac{\lambda_i^2\!+\!\lambda_j^2\!-\!2}{2} \right]\! \left(\textrm{i}\omega\tau\right)^n,
    \label{eq:elasticitytensor}
\end{equation}\normalsize

\noindent with ${I}_{jikl} = \left(\delta_{jk}\delta_{il}+\delta_{jl}\delta_{ik}\right)$, and $\lambda_i,\lambda_j$ the stretch ratios in directions $\ten{e_i}$ and $\ten{e_j}$.
Two distinct contributions show up here. The first one, $C^0_{jikl}(\lambda)$, are the components of a static tensor modified in order to take into account the effect of the static stretch according to the hyperelastic Mooney-Rivlin formalism (see Appendix~B for details).
The second term incorporates the effects of viscoelasticity~\cite{antman_2004,destrade_2009}, thus integrating the impact of frequency. It features a dependence on both the stretches $\lambda$ and the frequency $\omega$, and underlines the interdependence between these two variables. In fact, only if the second coefficient $\beta'$ vanishes, the elasticity tensor $\ten{C^\omega}$ given by equation~(\ref{eq:elasticitytensor}) is the sum of a function of $\lambda$ and of a function of $\omega$.
Instead, for our elastomer, a value of $\beta' = 0.29$ was determined in a previous work~\cite{delory_eml_2023}.

In this study, we are interested in displacements in a strip, meaning that boundary conditions need to be incorporated. Here, strip edges are free to move, thus the shear stress must vanish. After integrating these boundary conditions together with injecting a propagating solution for the displacement, non-zero solutions are find as zeros of a determinant. This yields to a dispersion relation that takes the form of a transcendental equation which can only be solved numerically.
Instead, here we employ a semi-analytical algorithm based on the Spectral Collocation Method~\cite{trefethen_spectral_2000,weideman_matlab_2000,adamou_spectral_2004,dubuc_2018,kiefer_elastodynamic_2022,kiefer_computing_2023} (SCM). The implementation of Ref.~\cite{kiefer_2022} dedicated to plates is adapted in order to consider the strip two-dimensional cross-section and is available in Ref.~\cite{kiefer_2023}. Assuming a periodic solution propagating along $x_1$ with wavenumber $k$, that is $\ten{u}(k, x_2, x_3, \omega) \e^{\iu (k x_1 - \omega t )}$, yields the following dispersion relation (see details in Appendix~C):

\begin{equation}
    \left[ (\iu k)^2 \ten{L}_2 + \iu k \ten{L}_1 + \ten{L}_0 + \omega^2 \ten{M} \right] \ten{u} = 0 \,.
    \label{eq:eigendiff}
\end{equation}

Therein, $\ten{L}_i$ and $\ten{M}$ are matrices of the discrete problem.
The above represents an algebraic eigenvalue problem for the eigenpair $(\omega^2,u)$ parameterized by $k$, as is common in commercial software. Alternatively, it can be solved for the eigenpair $(k, u)$ that is parameterized in $\omega$, which is particularly useful for frequency-dependent material parameters.
Choosing different values for $\omega$ and solving the quadratic eigenvalue problem with conventional methods yields the sought dispersion curves $k(\omega)$. Note that due to viscoelasticity, wavenumbers~$k$ are complex valued, while $\omega$ remains a real quantity. This is handled naturally by the eigenvalue solver and represents no difficulty.

The first two eigenvalues provide theoretical phase velocities $\omega/k$ depicted as lines in Figure~\ref{fig:LowFrequency}B and C.
The theoretical trends effectively coincide with all our measurements, without any fitting procedure (all the parameters correspond to Ref.~\cite{delory_eml_2023} devoted to an Ecoflex plate), within the investigated extension and frequency ranges.
For frequencies below 40~Hz, the wavelength of the compression mode approaches the dimensions of the plate. Consequently, the systematic extraction of the phase velocity becomes more sensitive to noise, leading to observable deviations of certain experimental points from the theoretical predictions.
The decomposition of incremental displacements, together with three-dimensional equations turn out to be the key to a proper description of strip dynamics. Note that our approach can be readily extended to a wide range of structures with different material rheology or elasticity laws.

%%%%%%%%%%%%%%%%%%%%%%%%%%%%%%%%%% CUT-OFF MODES %%%%%%%%%%%%%%%%%%%%%%%%%%%%%%%%%%%%%%%
\section*{Cut-off modes}

\begin{figure*}
    \centering
    \includegraphics[width=\textwidth]{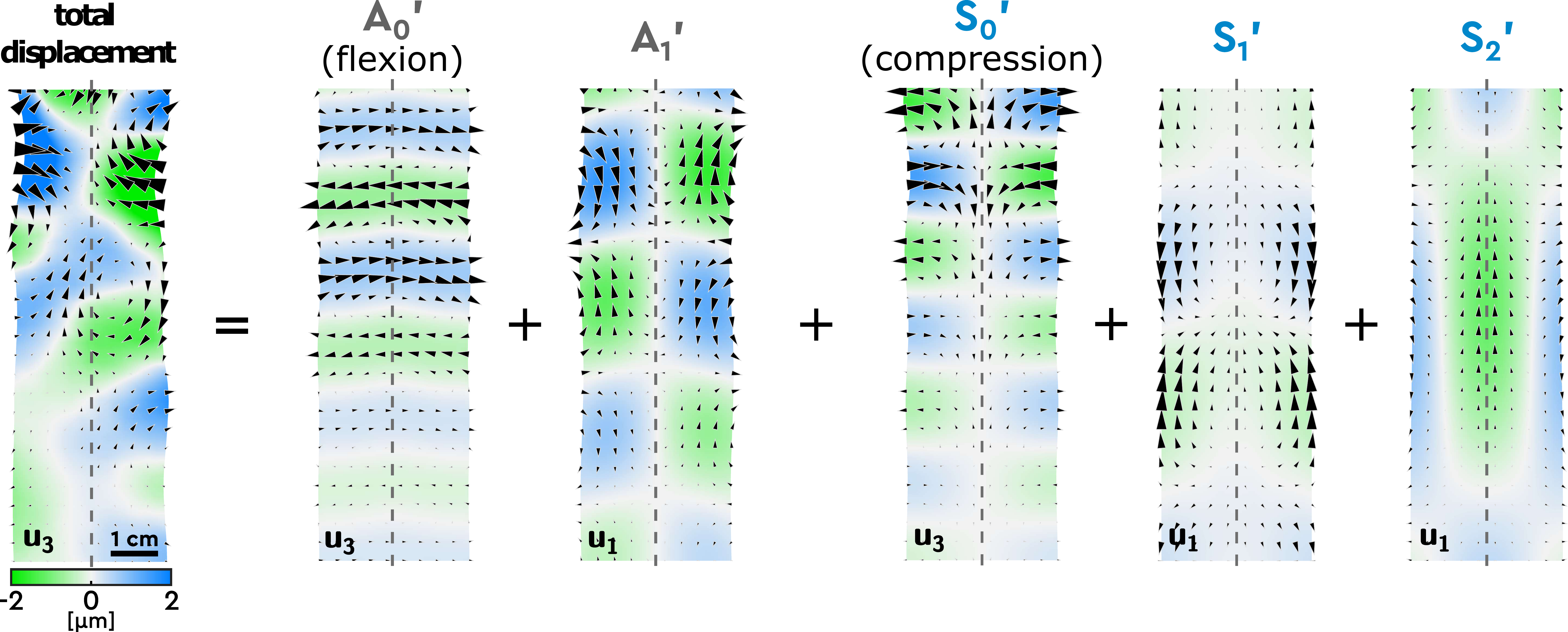}
     \caption{Field Decomposition at $f=200$~Hz --- Experimental in-plane total displacement field and eigenmode decomposition performed with a singular value decomposition algorithm. We display the fields with arrows, and the indicated displacement component with a colormap. Note that modes $A_0'$ and $S_0'$ correspond, respectively, to flexion and compression at lower frequencies, but both seem as transversely polarized waves at this frequency.}
    \label{fig:decomposition}
\end{figure*}

\subsection*{Free strip}
In the previous section, we observed that the strip operates as a finite waveguide. As a consequence, it has the capacity to host an infinite amount of eigenmodes. Up to this point, we have essentially discussed the behavior of two fundamental in-plane modes: compression and flexion. These two adequately account for the physics within the low-frequency range.
However, above certain cut-off frequencies, other contributions are likely to emerge. In other words, the displacement response must be projected onto a basis with additional eigenmodes. For instance, at 200~Hz, the field consists in the superposition of five eigenmodes (Figure~\ref{fig:decomposition}), which we separated thanks to a singular value decomposition~\cite{delory_JASA_2022}.
Here, modes are labelled according to a convention inspired by Lamb waves~\cite{laurent_JASA_2020,delory_JASA_2022}. Specifically, the label $S_n'$ (resp. $A_n'$) indicates a displacement field that is symmetric (resp. antisymmetric) along strip width (\textit{i.e.} across the dotted line in Figure~\ref{fig:decomposition}).
The index $n$ refers to the mode order, which is equivalent to the number of nodes in the transverse direction $\ten{e_3}$. These modes have been documented in prior studies~\cite{lanoy_PNAS_2020,delory_JASA_2022}, but their response to an external static stress remains unexplored. Note that the two fundamental modes ($n=0$) correspond at low frequencies to the compression ($S_0'$) and flexion ($A_0'$) described previously.

\begin{figure*}
    \centering
    \includegraphics[width=\textwidth]{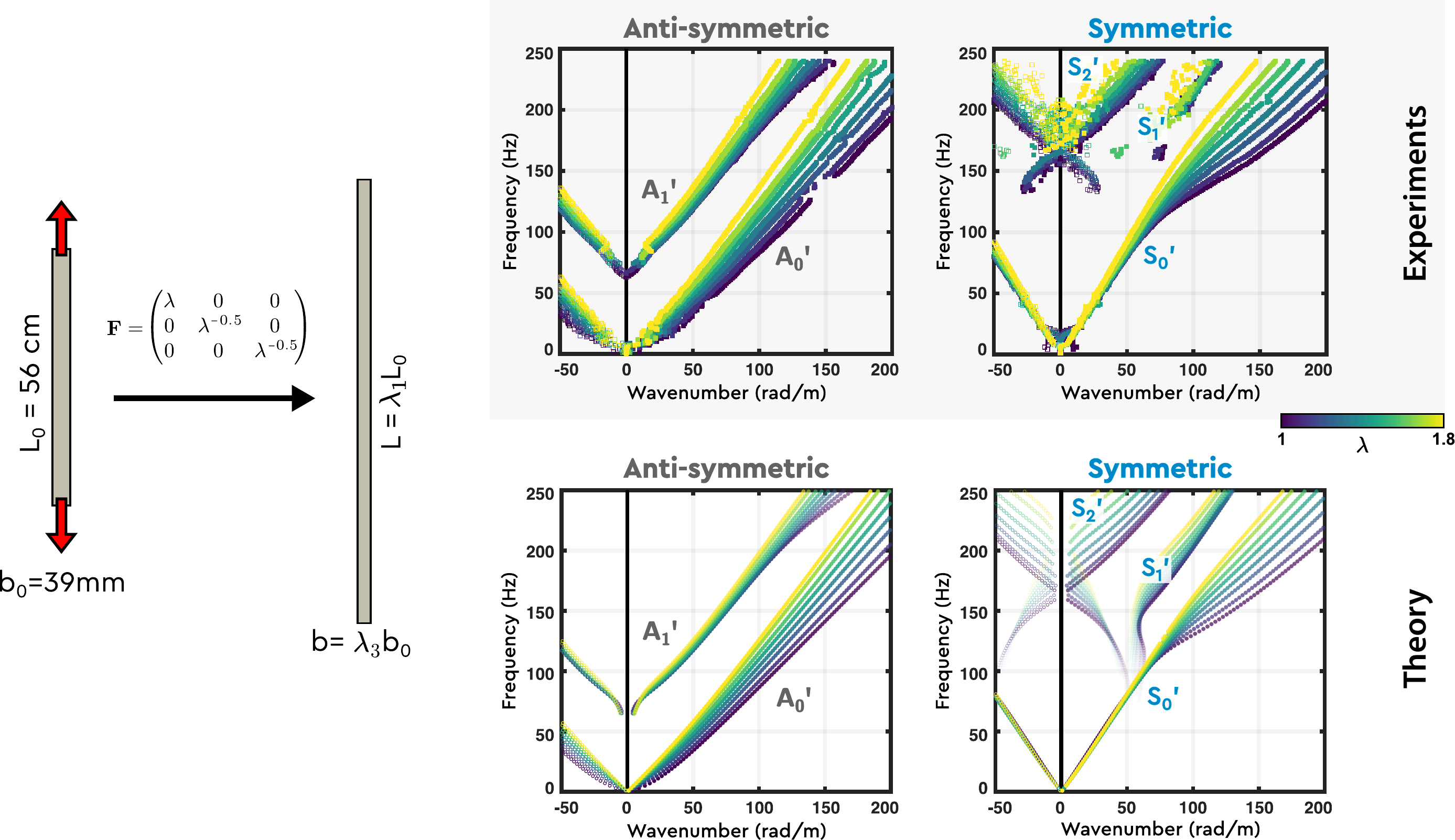}
    \caption{\textbf{Dispersion in a soft strip subjected to a uniaxial elongation ---} The deformation gradient $\mathbf{F}$ and the geometry are recalled for a free strip subjected to a uniaxial tension. Experimental dispersion curves of antisymmetric and symmetric modes in the elongated strip for several values of the stretch ratio $1\le \lambda \le 1.8$ are shown in the top line. The theoretical predictions (bottom line) are obtained using SCM (see text).}
    \label{fig:free}
\end{figure*}

Thanks to our setup, which provides reliable data up to 250~Hz, and empowered by our semi-analytical model, we can now comprehensively investigate the influence of an external stretch on the behavior of all modes, rather than restricting our analysis to just the two low-frequency modes. 
All experimental and numerical results are compiled in the dispersion diagram in Figure~\ref{fig:free}, which showcases the experimental points (top row) and the corresponding numerical outcomes (bottom row). These were acquired from the same soft strip subjected to varying amounts of longitudinal static stretch, ranging from $\lambda=1$ (depicted in blue) to $\lambda=1.8$ (depicted in yellow), and encompassing frequencies up to 250~Hz. For enhanced clarity, symmetric and anti-symmetric modes are presented in separate diagrams.

As one can notice, the theory effectively captures our experimental observations. As anticipated, only two modes exist at low frequencies ($A_0'$ being the flexural wave, and $S_0'$ the compressional wave). A third one ($A_1'$) emerges above 75~Hz, followed by two additional modes ($S_1'$ and $S_2'$) appearing roughly at 150~Hz.
Of particular interest is the uneven impact of longitudinal stretching on these branches. As emphasized in an earlier section, $S_0'$ (compression) seems nearly immune to it at low frequencies, a characteristic shared by $A_1'$. Conversely, the other modes exhibit greater sensitivity to stretching, notably in their slopes (\textit{i.e.} group velocity) but also, in the case of $S_1'$ and $S_2'$, in their cut-off frequencies.

A striking result is the change in the behavior of $S_0'$ with stretching when increasing frequency. Note how the branches spread out above 100~Hz. This feature provides a valuable hint towards understanding the governing mechanism. Indeed, at low frequencies $S_0'$ is essentially polarized in the longitudinal direction, as depicted in Figure~\ref{fig:LowFrequency}B, which is why it is commonly called the compressional mode. However, its dominant polarization switches as the frequency increases. On the displacement map acquired at 200~Hz (see Figure~\ref{fig:decomposition}), $S_0'$ indeed appears essentially polarized in the transverse direction. 

This strongly suggests that polarization is a determining criterion. This conclusion is further supported by the fact that, on one hand, both $A_0'$ and $S_2'$ are essentially polarized in the transverse direction (as depicted in Figure~\ref{fig:decomposition}) and turn out to be significantly influenced by the degree of stretching. On the other hand, $A_1'$ is characterized by a longitudinal polarization (as seen in Figure~\ref{fig:decomposition}) and proves resilient to stretching.
See more details in Appendix~D, where the dispersion curves in Figure\ref{fig:free} are displayed with a colormap rendering their polarization.

Finally, let us take a closer look at the $S_1'$ and $S_2'$ branches. See how their two cut-off frequencies coincide, resulting in the emergence of a crossing at $k=0$.
This degeneracy goes together with a linear dispersion for both the $S_1'$ and $S_2'$ branches. These features are signatures of a so-called Dirac cone. 
This kind of crossing, first evidenced in the context of electrons traveling in graphene~\cite{neto2009electronic} holds significant implications in diverse fields like semiconductor physics~\cite{ashcroft2022solid}, wave propagation control~\cite{Ochiai2009}, and communication systems~\cite{ozawa2019topological}. 
In the context of soft matter, this singularity was recently evidenced~\cite{lanoy_PNAS_2020} and pertains to incompressible materials ($\nu=1/2$) for the strip geometry~\cite{laurent_JASA_2020}.
Here, we show that the Dirac cone is relatively robust to extensional stress. Our measurements also demonstrate that the frequency of occurrence can be controlled by adjusting the amount of static stretch, which can be relevant when considering technological developments.

\begin{figure}
    \centering
    \includegraphics[width=.49\textwidth]{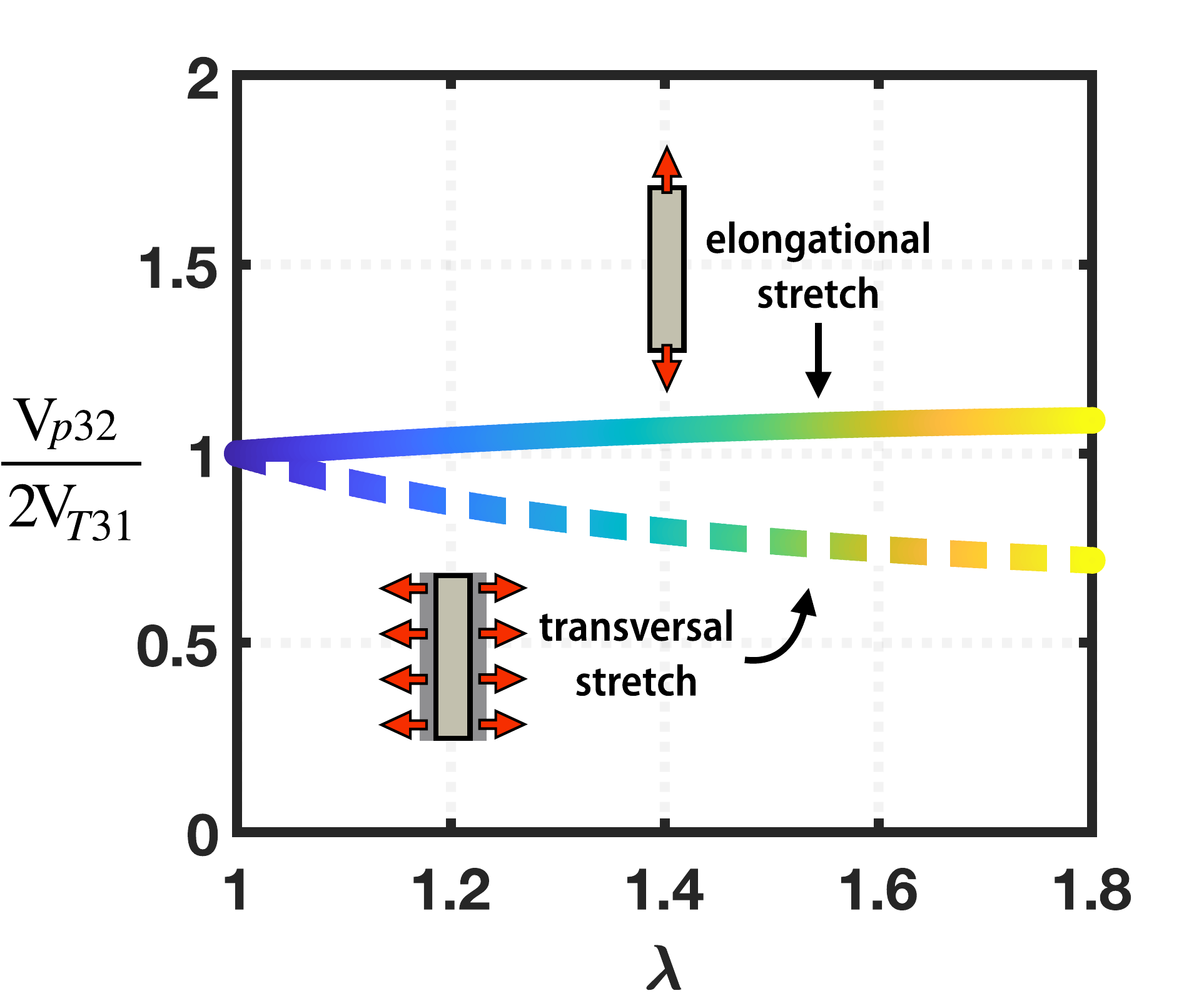}
     \caption{Cut-off frequencies --- Evolution of the ratio of phase velocities $V_{p32}/2V_{T31}$ with the applied deformation in the case of an elongational stretch (solid line) and of a transversal stretch (dashed line).}
    \label{fig:cutoff_freq}
\end{figure}

To understand these two effects, it is essential to go back to the expressions of the cut-off frequencies. As shown in a recent contribution building a plate-strip equivalence~\cite{laurent_JASA_2020}, one can demonstrate that the cut-off is governed by a coupling between the two following modes: a shear wave propagating at $V_T=\sqrt{E/3\rho}$ and a {\it plate} wave propagating at $V_p=2V_T$. Their cut-off frequencies are multiples of $V_T/2b$ and $V_p/2b$~\footnote{These are obtained by considering the cut-off frequency of the Lamb modes in a plate (\textit{i.e.} $ n V_T/2h$ and $m V_L/2h$) and replacing (i) the velocity of longitudinal waves $V_L$ by the velocity of the {\it plate} wave $V_p$ as well as (ii) the thickness of the plate $h$ by the width of the strip $b$.
The equivalence has been shown to be accurate provided that the strip is sufficiently thin ($b\gg h$) and for frequencies below the first thickness cut-off frequency ($V_T/2h$, which is here of $\sim$1~kHz).}.
Of course, the directions associated with theses velocities should be specified since the material becomes anisotropic upon stretching~\cite{delory_eml_2023}. Here, the relevant transverse wave responsible for the cut-off frequency propagates in the $\ten{e_3}$ direction and is polarized in $\ten{e_1}$ direction. Its phase velocity writes~\cite{rogerson_1995,delory_eml_2023} $V_{T31}=\omega / \text{Re}\left[k_{T31}\right]$ with $k_{T31}/\omega = \left[C^{\omega}_{3113}/\rho\right]^{-1/2}$.
Conversely, the {\it plate} wave involved for the cut-off frequency propagates again in $\ten{e_3}$ direction but is also polarized in $\ten{e_3}$ direction. Since it has also a displacement in $\ten{e_2}$ direction, its velocity is actually completely governed by moduli implying these two directions. It can be expressed~\cite{rogerson_1995,delory_eml_2023} with $k_{p32}/\omega=\left[C^{\omega}_{3223} + 3 C^{\omega}_{2332})/\rho\right]^{-1/2}$.
For the Dirac cone to exist, the first ($V_{p32}/2b$) and second ($V_{T31}/b$) non-zero cut-off frequencies should coincide. Figure~\ref{fig:cutoff_freq} reports the evolution of the ratio between these two quantities as a function of the stretch ratio $\lambda$ (solid line). For this kind of deformation, it appears that the ratio remains relatively close to 1 (starting from exactly 1 at $\lambda=1$ and plateauing at 1.08 for $\lambda=1.8$), which explains why the cone seems present regardless of the applied tension.
This relative robustness can also be harnessed when looking at the polarization of the field near the Dirac cone (Figure~\ref{fig:polarisation} in Appendix~D): as a consequence of the degeneracy between two modes with orthogonal polarizations the dispersion curve near this point exhibits the presence of the two polarizations. Nevertheless, one has to keep in mind that in solid state physics a Dirac cone is a true degeneracy~\cite{neto2009electronic,ashcroft2022solid}, unlike here the two branches do not really cross in the complex wavenumber domain since they are associated to different imaginary parts~\footnote{
By also studying the wavenumber imaginary part, and not only its real part, it is more like two straight lines that approach but avoid each other by passing through the wavenumber complex plane, and not just the real wavenumber axis, which is rendered by the transparency in Figure~\ref{fig:free}. When the cut-off frequencies are close but not exactly the same, the two linear branches avoid each other more, still without crossing, and it looks as if the Dirac cone still exists. This increase in the imaginary part of the wavenumber with the stretch ratio around the Dirac cone makes it even harder to measure it experimentally.}.

\begin{figure*}
    \centering
    \includegraphics[width=\textwidth]{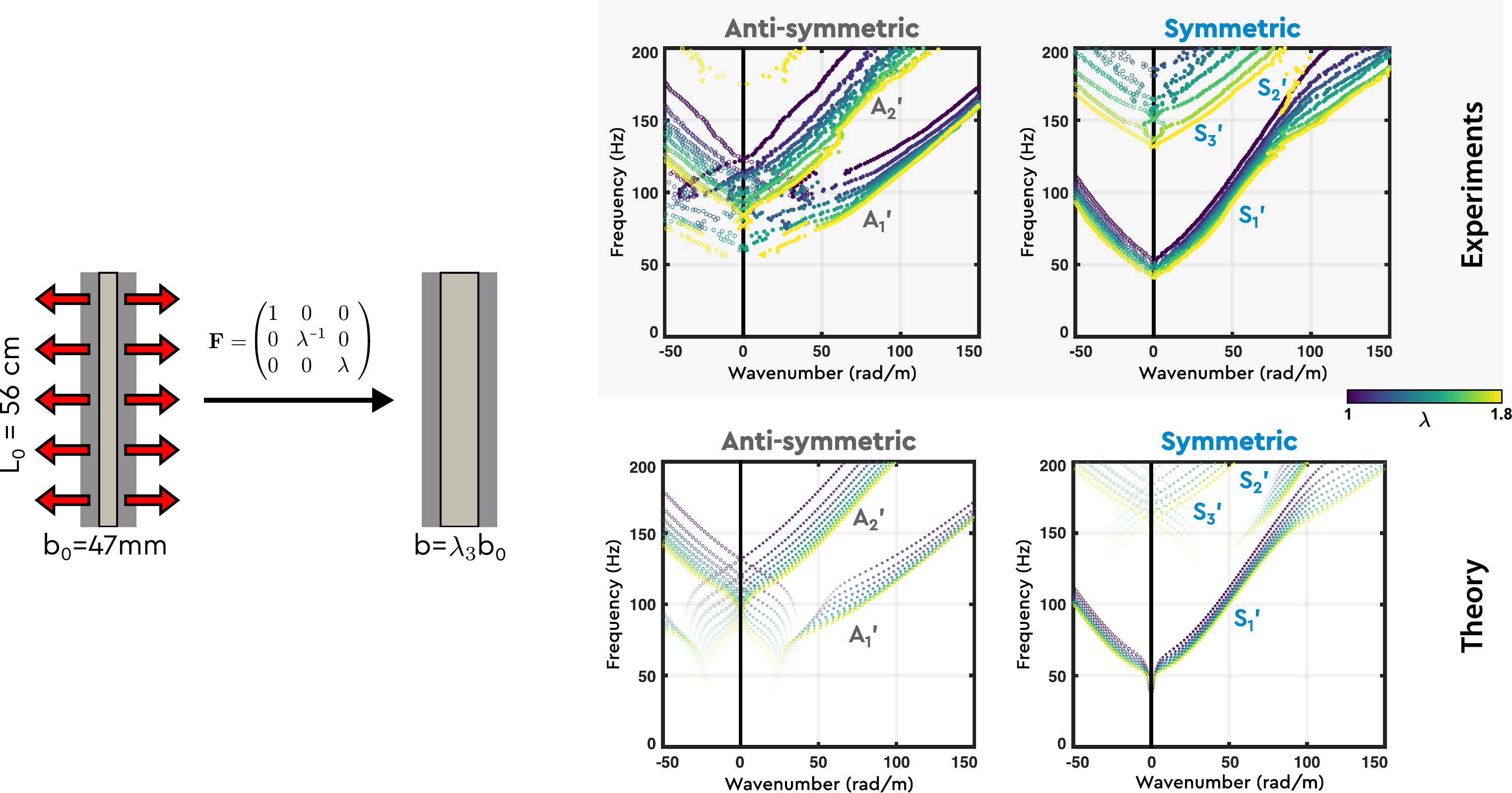}
    \caption{\textbf{Dispersion in a soft strip, clamped at the edges, subjected to a planar elongation ---} (left) The deformation gradient $\mathbf{F}$ and the geometry are recalled. Experimental dispersion curves of antisymmetric and symmetric modes for several values of the stretch ratio $1\le \lambda \le 1.8$ (top line). The theoretical predictions (bottom line) are obtained using SCM (see text).}
    \label{fig:fixed}
\end{figure*}

\subsection*{Clamped strip}

We now consider an alternative configuration which also supports the existence of a Dirac cone~\cite{lanoy_PNAS_2020}, wherein a soft strip is clamped at its lateral edges. By adjusting the distance between the clamps, we are able to induce a transverse stretch (as illustrated in the inset of Figure~\ref{fig:cutoff_freq} and in Figure~\ref{fig:fixed}). This new deformation prompts alterations in both the elasticity tensor and the governing velocities, such as $V_{T31}$ and $V_{P32}$.
In this configuration, which corresponds to a planar elongation, we obtain that the first and second non-zero cut-off frequencies are gradually moving away as the stretching increases (as pictured by the dashed line in Figure~\ref{fig:cutoff_freq}). As a result, we anticipate a clear disruption of the Dirac cone. This can be verified thanks to the dispersion curves presented in Figure~\ref{fig:fixed}. Once again, theory effectively matches experimental data. 

It is worth pointing out that, in this case, the cone pertains to anti-symmetric modes. The linear crossing, which is its signature, occurs in the initial state (blue symbols), but splits in two separate branches as $\lambda$ is increased (yellow symbols). This time, cut-off frequencies no longer coincide, and instead we see two branches that plunge completely into the complex plane.
This trend is apparent in the experimental curve and becomes more evident in theoretical plots. This is particularly visible in Figure~\ref{fig:polarisation} in Appendix~D where the mixed polarization (orange) for the unstretched case gives rise to two different branches with clear orthogonal polarizations (yellow and red) for stretched cases. In addition, one might notice the absence of the two fundamental modes in this configuration. This is directly linked to the clamping process, which suppresses rigid body motions.

This configuration provides a demonstration that the three-dimensional model can readily be extended to various sets of boundary conditions and to different kinds of static deformations. Also, this illustrates how applying an external static stretch serves as a mean to tailor dispersion, and consequently, tune the overall strip dynamics. 
It provides a compelling illustration of the possibility to control the structure response with an adequate static deformation. This paves the way towards the design of soft tunable structures.

%%%%%%%%%%%%%%%%%%%%%%%%%    CONCLUSION   %%%%%%%%%%%%%%%%%%%%%%%%%%%%
\section*{Conclusion}

This article investigates in-plane dynamics of a soft strip experiencing a significant static deformation in the longitudinal or transverse directions. To that end, our approach consists in monitoring the propagation of elastic waves within the strip. Our experiments, supported by a semi-analytical model, reveal that static stretching strongly impacts the dynamics of the strip.
Interestingly, we observe that certain vibration modes seem nearly immune to the external stretching, while others display a high sensitivity. We find that this sensitivity is essentially governed by the displacement's polarization. 

These findings are well explained by incorporating both the rheology and hyperelastic constitutive law of the material, giving a total of 5 parameters. We first exemplified this by modelling the effect of stretching on the compression and flexion of the strips with simple one dimensional models. Then, to render the whole waveguiding phenomenon occurring at higher frequencies, we called on a generalized three-dimensional formalism.
Within the framework of incremental displacement theory, which involves a small dynamic perturbation superimposed on a significant static deformation, we managed to derive the full dynamic response.

In biological tissues and organic matter, flexible structures under tension play a pivotal role in various physiological processes. Our research represents a step towards a better understanding of the mechanics of vocal folds, tendons and muscles among others. Furthermore, we demonstrate that observing wave propagation through a simple strip provides concrete insights into rheological properties and stress state of the material. These findings set the stage for refining ultrasound elastography techniques~\cite{gennisson_2007,crutison_2022}, which currently lack quantitative capabilities for imaging stretched organs.

Conversely, because we can make trustworthy predictions, the external deformation is no longer an obstacle. Instead, it becomes a valuable tool for shaping the response and fine-tuning overall structure dynamics.
These results unlock interesting perspectives in terms of design of adaptive soft structures, with potential applications in the fields of vibration mitigation, energy harvesting and soft robotics. 

\section*{Acknowledgements}
This work has received support under the program « Investissements d’Avenir » launched by the French Government and partially by the Simons Foundation/Collaboration on Symmetry-
Driven Extreme Wave Phenomena. A.D. acknowledges funding from French Direction G{\'e}n{\'e}rale de l'Armement.

%%% REFERENCES %%%
%\bibliographystyle{apsrev4-2.bst}
%\bibliography{monbib_arXiv.bib}


\newpage

%%%%%%%%%%%%%%%%%%%    SUPPLEMENTARY    %%%%%%%%%%%%%
\section*{Appendices}
%%%%%%%%%%%%%%%%%%%%%%%%%%%%%%%% Mécanique d'une poutre 
\subsection*{Appendix A: Derivation of the Mooney-Rivlin equivalent Young's modulus}

In the realm of continuum mechanics, the Mooney–Rivlin constitutive law~\cite{mooney1940theory,rivlin1948large,beatty2001seven,puglisi_2016,destrade_2017} serves as a hyperelastic model to account for the deviation from Hooke's law for large deformations. It is defined by its strain energy density function $W$. This function is constructed as a linear combination of the first two invariants $I_1$ and $I_2$ of the left Cauchy–Green deformation tensor ($\mathbf{B}$), defined in Appendix~B. For an incompressible Mooney–Rivlin material, the strain energy density function takes on the following expression: 

\begin{equation}
    W = C_1 (I_1-3) + C_2 (I_2-3).   
\end{equation}

\noindent This model requires 2 constants to describe the mechanical properties. As Hooke's law already provides a scalar quantity, namely the Young's modulus $E_0$, we prefer to rewrite this energy function as:

\begin{equation}
 W=\frac{E_0}{6}\left[(1-\alpha)(I_1-3)+\alpha(I_2-3)\right].
\end{equation}

\noindent Note that $\alpha=0$ (or equivalently $C_2=0$) corresponds to a Neo-Hookean solid~\cite{ogden_1997,puglisi_2016,destrade_2017} which only accounts for geometrical non-linearities.

In the specific scenario of a uniaxial elongation along direction $\ten{e_1}$, we can define a stretch ratio $\lambda_1 = \lambda$. For an incompressible material, stretch ratios in other directions $\ten{e_2}$ and $\ten{e_3}$ simply write $\lambda_2=\lambda_3 = 1/\sqrt{\lambda}$ since volume has to be conserved. Then, it becomes possible to calculate the true stress (Cauchy stress) as equation~(\ref{eq:MooneyRivlin}) in the main text. This is done by evaluating the following equation which arises from the condition of vanishing lateral tractions: $$ \sigma = \lambda_1~\frac{\partial W}{\partial \lambda_1} - \lambda_3~\frac{\partial W}{\partial \lambda_3}. $$

This expression clearly evidences that the true stress does not linearly grow with elongation. Another way to interpret this formula would be to keep using Hooke's law, but considering a Young's modulus that depends on the elongation. By doing so, we implicitly consider the undeformed configuration as a reference. Then, we need to evaluate the engineering (or nominal) stress which relates internal forces in the deformed configuration with areas from the reference. For the incompressible material it simply gives:

\begin{equation}
    \sigma^{\textrm{eng}}=\sigma \lambda_2\lambda_3=\frac{\sigma}{\lambda}.
\end{equation}

Finally, one easily extracts the desired elongation-dependent Young's modulus~\cite{zhao2021elastic} of equation~(\ref{eq:Young_MooneyRivlin}) as: 

\begin{align}
    E(\lambda)&=\frac{\textrm{d}}{\textrm{d}\lambda}\left(\frac{\sigma}{\lambda}\right) \nonumber \\
    &=\frac{E_0}{3}\left[(1-\alpha) \left(1 + \frac{2}{\lambda^3}\right) + \frac{3\alpha}{\lambda^4}\right].
\end{align}

%%%%%%%%%%%%%%%%%%%%%%%%%%%%%%%%%%%%%%%%%%%%%%%%%%%%%%%%%%%%%%%%%%%%%%
%%%%%%%%%                Acousto elasticity theory
%%%%%%%%%%%%%%%%%%%%%%%%%%%%%%%%%%%%%%%%%%%%%%%%%%%%%%%%%%%%%%%%%%%%%%
\subsection*{Appendix B: Coefficients of the elasticity tensor $\ten{C}^0\left(\lambda\right)$ for incremental waves in a hyperelastic solid}

A compressible version of the Mooney-Rivlin model was actually used in this work, with a bulk modulus of 1~GPa, as an input for the Spectral Collocation Method (SCM), described in Appendix~C. Thus, we switch to compressible formulations for the quantities of interest. Note that this does not change any of the above conclusions since the effect of compressibility for this set of parameters is negligible during a tensile test.

The hyperelastic constitutive law relies on the use of a strain energy density function~$W$ which contains all mechanical properties. The associated stress $\boldsymbol{\sigma}$ is given by:

\begin{equation}
    \boldsymbol{\sigma} = \frac{1}{J}\mathbf{F}\cdot\frac{\partial W}{\partial \mathbf{E}}\cdot\mathbf{F}^{{\rm T}} \,\text{ with } \mathbf{E}\!=\! \frac{\mathbf{F}^{\mathrm{T}}\!\cdot\!\mathbf{F}\!-\!\mathbf{1}}{2} \text{ and } J\!=\!\text{det}\left(\mathbf{F}\right),
    \label{eq:hyperelasticlaw}
\end{equation}

where $\mathbf{F} = \mathbf{1}+\nabla\mathbf{u}$ is the deformation gradient, $\mathbf{u}=\mathbf{x}-\mathbf{X}$ is the displacement, $\mathbf{1}$ is the second-order identity tensor, and $\mathbf{E}$ is the Green Lagrangian strain tensor, as introduced in Refs.~\cite{ogden_1997,destrade_2007}.
For an isotropic solid, $W$ is a function of principal invariants of the left ($\mathbf{B}=\mathbf{F}\cdot \mathbf{F}^{{\rm T}}$) Cauchy--Green tensor:

\begin{align*}
    I_1 &= \text{Tr}\left(\mathbf{B}\right) = \lambda_1^2 + \lambda_2^2 + \lambda_3^2\\
    I_2 &= \frac{1}{2}\left(\text{Tr}\left(\mathbf{B}\right)^2-\text{Tr}\left(\mathbf{B}^2\right)\right) = \lambda_2^2\lambda_3^2 + \lambda_1^2\lambda_3^2 + \lambda_1^2\lambda_2^2 \\
    I_3 &= \text{det}\left(\mathbf{B}\right) = \lambda_1^2\lambda_2^2\lambda_3^2 = J^2
\end{align*}

This ensures invariance of~$W$ under a permutation of $\left(\lambda_1,\lambda_2,\lambda_3\right)$.
The strain energy density function~$W$ for a Mooney-Rivlin hyperelastic model writes:
\begin{equation}
    W = \frac{E_0}{3} \left[ (1-\alpha)\left(\frac{I_1}{J^{2/3}}-3\right) + \alpha\left(\frac{I_2}{J^{4/3}}-3\right) \right] + \frac{\kappa}{2}\left(J-1\right)^2
\end{equation}
with the bulk modulus $\kappa \gg E_0$.
From this, an incremental approach is built to describe waves in a prestressed body, and equation~(\ref{eq:elasticitytensor}) is obtained but with replacing the elasticity tensor by $\ten{C}^0$, with its coefficients expressed as:
\begin{align}
    C^0_{iijj} &= \frac{\lambda_i\lambda_j}{J} \, W_{ij},& \nonumber\\
    C^0_{ijij} &= \frac{\lambda_i\lambda_j}{J} \, \frac{\lambda_j W_i - \lambda_i W_j}{\lambda_i^2-\lambda_j^2} & \text{if } (i\neq j, \lambda_i\neq\lambda_j), \nonumber\\
    C^0_{ijij} &= \frac{\lambda_i^2 W_{ii}-\lambda_i\lambda_j W_{ij}-\lambda_i W_i}{2J} & \text{if } (i\neq j, \lambda_i=\lambda_j),\label{eq:elastictensor}\\
    C^0_{ijji} &= \frac{\lambda_i^2}{J} \, \frac{\lambda_i W_i - \lambda_j W_j}{\lambda_i^2-\lambda_j^2} & \text{if } (i\neq j, \lambda_i\neq\lambda_j), \nonumber\\
    C^0_{ijji} &= \frac{\lambda_i^2 W_{ii}-\lambda_i\lambda_j W_{ij}+\lambda_i W_i}{2J} &  \text{if } (i\neq j, \lambda_i=\lambda_j),\nonumber
\end{align}
where $W_i = \dfrac{\partial W}{\partial \lambda_i}$ and $W_{ij} = \dfrac{\partial^2 W}{\partial \lambda_i \partial \lambda_j}$.\\
Here, expressions are slightly different from the ones commonly found in the literature~\cite{ogden_1997,destrade_2007} because dot and double-dot product conventions are different. The elasticity tensors of these two formulations are related by a simple permutation of the last two indices.

%% %%%%%%%%%%%%%%%%%%%%%%%%%%%%%     SCM    %%%%%%%%%%%%%%%%%%%%%%%
\subsection*{Appendix C: Computing guided waves in the strip with SCM}

The computational method\textcolor{red}{~\cite{kiefer_2023}} consists of three fundamental steps: (i) derive the boundary-value problem that describes plane guided waves, (ii) replace the differential operators by spectral differentiation matrices to obtain a discrete approximation of the guided wave problem, and (iii) use standard numerical methods to solve the resulting algebraic eigenvalue problem.

Step (i) consists of inserting the plane wave ansatz for the displacements into the equation of motion given in (\ref{eq:equation_of_motion_underformed}). After re-arranging the terms this yields (in symbolic tensor notation):

\begin{align}
    & \left[ (\iu k)^2 \ten{c}_{11} + \iu k (\ten{c}_{21} + \ten{c}_{12}) \partial_2 + \iu k (\ten{c}_{31} + \ten{c}_{13}) \partial_3 + \ten{c}_{22} \partial_2^2 \right. \nonumber\\
    & \, \left. + (\ten{c}_{32} + \ten{c}_{23}) \partial_3 \partial_2 + \ten{c}_{33} \partial_3^2 + \omega^2 \rho \ten{1} \right] \cdot \ten{u} = \ten{0} \quad\text{on}\quad \Omega \,,
    \label{eq:motion_guided_waves}
\end{align}
where we have defined the second order tensors $\ten{c}_{ij} := \dirvec{i} \cdot \ten{C} \cdot \dirvec{j}$ with $i, j \in \{1, 2, 3\}$. A more detailed derivation for a plate can be found in Ref.~\cite{kiefer_computing_2023}.

Boundary conditions are needed in addition to (\ref{eq:motion_guided_waves}). The boundary~$\partial\Omega$ splits into one region~$\partial\Omega_\mup{N}$ where the strip is free (homogeneous Neumann boundary condition) and one region~$\partial\Omega_\mup{D}$ where it is clamped (homogeneous Dirichlet boundary condition). Writing $\dirvec{n}$ for the unit normal to the strip cross-section, \textit{i.e.} either $\dirvec{2}$ or $\dirvec{3}$, the homogeneous Neumann boundary condition reads:

\begin{equation}\label{eq:BC_Neumann}
    \dirvec{n} \cdot \ten{C} : \nabla \ten{u}
    = \left[\textrm{i}k \ten{c}_{n1} + \ten{c}_{n2}\partial_2 + \ten{c}_{n3}\partial_3 \right] \cdot \ten{u}
    = \ten{0} \quad \text{on} \quad \partial\Omega_\mup{N} \,.
\end{equation}

The clamped boundary condition, on the other hand, simply reads:

\begin{equation}\label{eq:BC_Dirichlet}
    \ten{u} = \ten{0} \quad \text{on} \quad \partial\Omega_\mup{D} \,.
\end{equation}

The equation of motion~(\ref{eq:motion_guided_waves}) together with the boundary conditions~(\ref{eq:BC_Neumann}) and~(\ref{eq:BC_Dirichlet}) constitute the boundary-value problem that describes guided waves in the strip.
Note that for a given value of $\omega$, it constitutes a quadratic differential eigenvalue problem with eigenvalue $k$ and eigenfunction $\ten{u}(x_2,x_3)$. Prestress and viscoelasticity are considered using the appropriate elasticity tensor given by $\ten{C}^{\omega}$ in equation~(\ref{eq:elasticitytensor}).
Hence, the stretched strip is processed in the same way as without prestress, except that the dimensions of the cross-section (width~$b$ and thickness~$h$) are iteratively adapted to the static pre-deformation currently being considered.

The discretization is performed in step (ii). To this end, the domain $\Omega = [0,h]\times[0, b]$ is discretized as suggested by Weideman and Reddy~\cite{weideman_matlab_2000} using Chebyshev spectral collocation. The first and second order differentiation matrices $D^{(10)}$ and $D^{(20)}$ of size $N \times N$ along the $x_2$-coordinate are computed using DMSUITE~\cite{weideman_matlab_2000}. We proceed similarly for differentiation along the $x_3$-coordinate, yielding matrices $D^{(01)}$ and $D^{(02)}$ of size $P \times P$. Next, the differentiation matrices in the ($x_2,x_3$)-plane are obtained as Kronecker products, denoted by ``$\otimes$'', between the former one-dimensional differentiation matrices. 
Concretely, this yields the matrices:
\begin{align}
    D_{23} &= D^{(01)} \otimes D^{(10)} \,,  & D_{2} &= I_P \otimes D^{(10)} \,, & D_{22} &= I_P \otimes D^{(20)}\,, \nonumber\\
    D_{3} &= D^{(01)} \otimes I_N \,,  & D_{33} &= D^{(02)} \otimes I_N \,,  & I_\mup{d} &= I_P \otimes I_N \,,
    \label{eq:diff_mats}
\end{align}

\noindent where $I_Q$ denotes the identity matrix of size $Q \times Q$.

Next, the partial derivatives in~(\ref{eq:motion_guided_waves}) as well as~(\ref{eq:BC_Neumann}) are replaced by the differentiation matrices given in~(\ref{eq:diff_mats}). When doing so, the multiplication of the differentiation matrices with the second order constitutive tensors $\ten{c}_{ij}$ needs to be interpreted again as Kronecker products. This finally yields:

\begin{align}\label{eq:motion_guided_waves_discr}
    & \left[ (\iu k)^2 \ten{c}_{11} \otimes I_\mup{d} + \iu k (\ten{c}_{21} + \ten{c}_{12}) \otimes D_2 + \iu k (\ten{c}_{31} + \ten{c}_{13}) \otimes D_3 \right. \nonumber\\
    & \quad \left. + \ten{c}_{22} \otimes D_{22} + (\ten{c}_{32} + \ten{c}_{23}) \otimes D_{23} \right.\\
    & \qquad \left. + \ten{c}_{33} \otimes D_{33} + \omega^2 \rho \ten{1} \otimes I_\mup{d} \right] u = 0 \,, \nonumber
\end{align}

\noindent where $u$ denotes the $3NP \times 1$ vector of $u_1, u_2, u_3$ displacements at the $NP$ collocation points. Hence, equation~(\ref{eq:motion_guided_waves_discr}) represents a linear system of size $3NP \times 3NP$.

The discrete boundary conditions are obtained similarly. The Neumann boundary condition from~(\ref{eq:BC_Neumann}) becomes:

\begin{equation}\label{eq:BC_Neumann_discr}
    \left[ \iu k \ten{c}_{n1} \otimes I_\mup{d} + \ten{c}_{n2} \otimes D_2 + \ten{c}_{n3} \otimes D_3 \right] u = 0 \,,
\end{equation}

\noindent while the fixed condition from~(\ref{eq:BC_Dirichlet}) is:

\begin{equation}\label{eq:BC_Dirichlet_discr}
    \ten{1} \otimes I_\mup{d} \, u = 0 \,.
\end{equation}

\begin{figure}
    \centering
    \includegraphics[width=7cm]{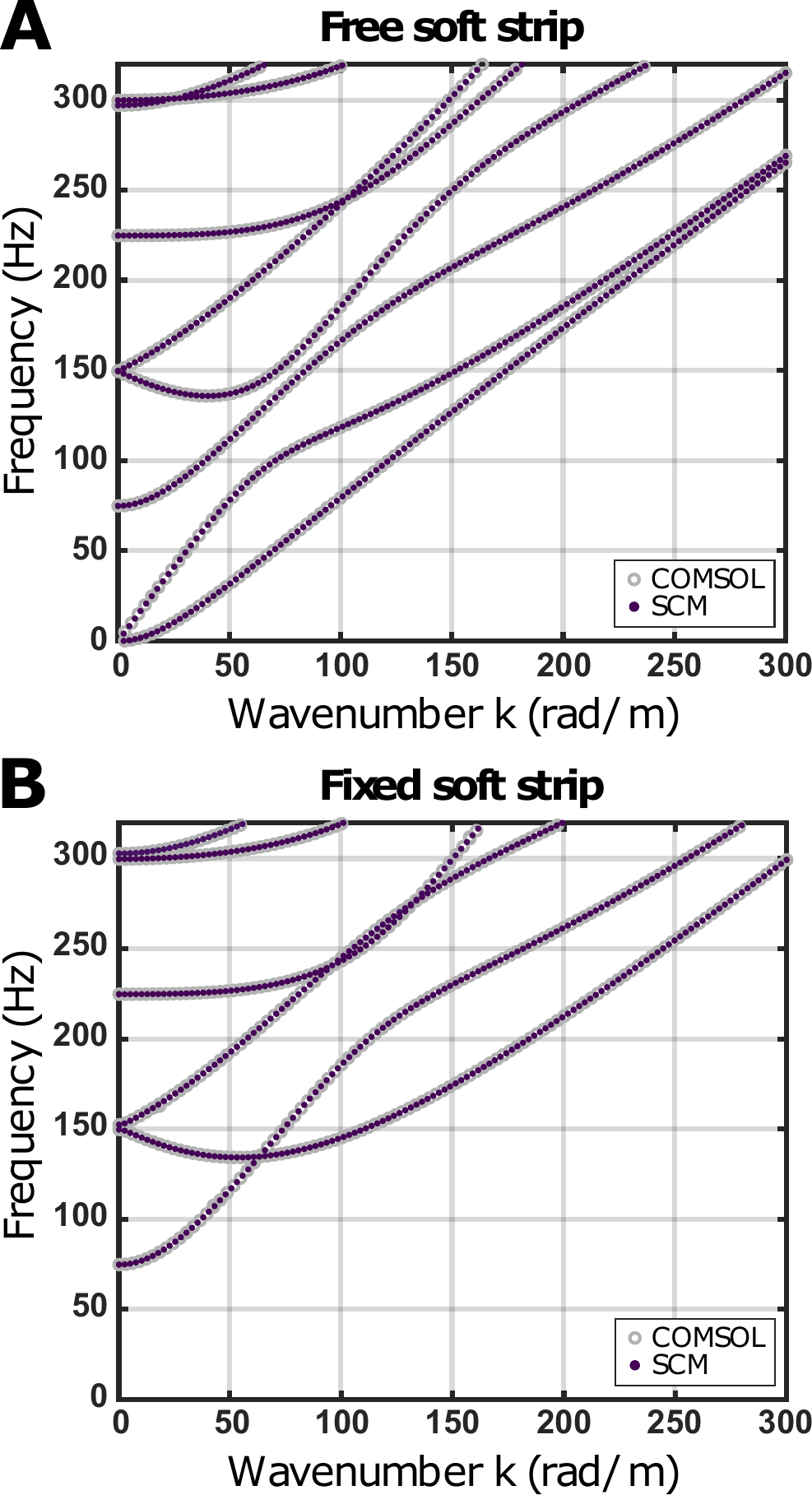}
    \caption{\textbf{Comparison of the SCM with solutions obtained with COMSOL Multiphysics ---} Numerically computed dispersion curves for in-plane guided waves in an undeformed soft elastic strip of size $h = \text{3\,mm} \times b = \text{40\,mm}$. Solutions coincide both for a free strip in (A) and for the partially clamped strip in (B).}
    \label{fig:scm_vs_comsol}
\end{figure}

Lastly, the boundary conditions need to be incorporated into (\ref{eq:motion_guided_waves_discr}). This is done by replacing the corresponding rows of (\ref{eq:motion_guided_waves_discr}) with the ones from (\ref{eq:BC_Neumann_discr}) or (\ref{eq:BC_Dirichlet_discr}), as appropriate. Denoting the final matrices with the mentioned replacements as $L_2$, $L_1$, $L_0$ and $M$, this finally leads to the equation~(\ref{eq:eigendiff}) in the main text. 

In order to test the implementation, we have considered a purely elastic material without prestress by setting $\ten{C}^{\omega} = \ten{C}$, and we present the results for in-plane guided waves in Figure~\ref{fig:scm_vs_comsol} for the two types of boundary conditions. For the sake of completeness, results are compared to the ones obtained by finite element method, with the commercial software COMSOL, showing a perfect agreement.

While the free strip computes seamlessly, the partially fixed strip leads to numerical difficulties due to the singularities present at the corners of the rectangular cross-section. 
Choosing different boundary conditions at the corners leads to somewhat different behaviors of the solutions. In any case, the singularities become less important when increasing $N$, $P$ and all solutions converge slowly towards the COMSOL Multiphysics reference.
We notice, however, that choosing $N$ even (odd) leads to a good representation of the mostly in-plane (out-of-plane) polarized waves. As we are only interested in the in-plane guided waves, we can choose $N$ and $P$ as before and obtain converged results for these waves. The described problem could be overcome by utilizing a finite element discretization instead.
As we are able to obtain accurate solutions for the waves of interest, we have stuck to the very fast Spectral Collocation Method to perform parametric studies of prestressed viscoelastic strips in this work. 

%%%%%%%%%%%%%%%%%  Polarization colorbar
\subsection*{Appendix D: Mode polarization and dependence on the deformation}

\begin{figure*}
    \centering
    \includegraphics[width=.8\textwidth]{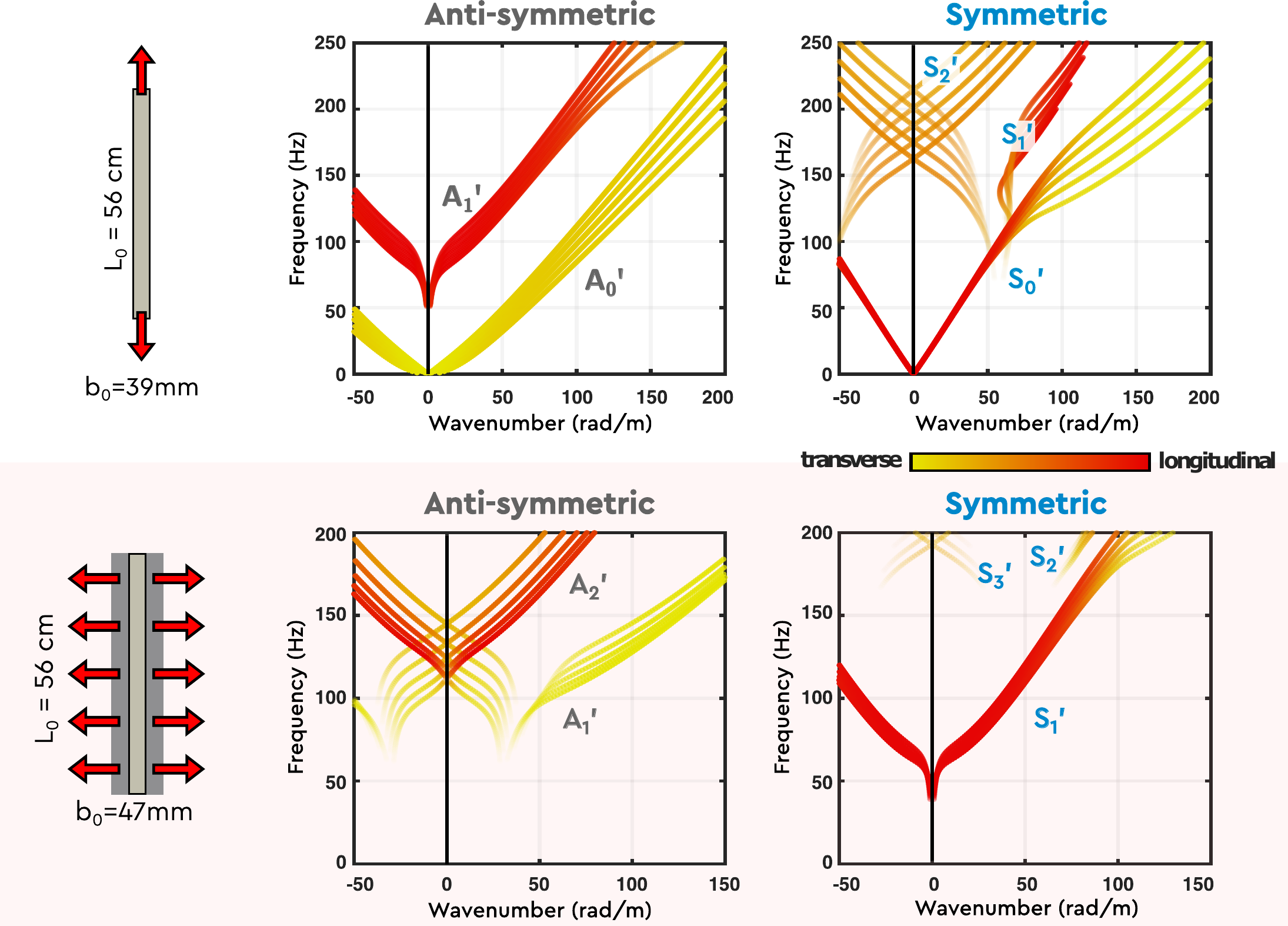}
     \caption{\textbf{Role of the polarization ---} Theoretical dispersion curves for the free strip under elongational stretch (top row) and the clamped strip under lateral stretch (bottom row). The color indicates the dominant polarization (yellow = transverse, red = longitudinal).}
    \label{fig:polarisation}
\end{figure*}

Dispersion curves of in-plane eigenmodes in stretched strips are redisplayed in Figure~\ref{fig:polarisation}, for various stretch ratios and configurations. In the top part, a free strip is subjected to a uniaxial stress, while in the bottom part, a strip with fixed edges is subjected to a planar stress. This time, color encodes for the mode polarization.

In practice, the spectral collocation method is used to solve equation~(\ref{eq:eigendiff}) at a fixed frequency $\omega$ to obtain the eigenpair $(k, \ten{u})$.
By studying $\ten{u}$, one can discriminate in-plane eigenmodes (from their out-of-plane counterparts) and their corresponding symmetry. One can also study their main polarization by evaluating mean values of $|u_1|^2$ and $|u_3|^2$ over the cross-section and compute the inverse tangent of the ratio $\int|u_1|^2/\int|u_3|^2$.
Then, it is possible to see whether the mode is mostly polarized in the $\ten{e_1}$ direction (longitudinal in red) or in the $\ten{e_3}$ direction (transverse in yellow).

The first and most obvious observation is the $S_0'$ mode polarization in a free strip. At low frequency, it appears red and all curves are superimposed. This is the so-called compressional mode discussed earlier and depends very little on the applied prestress, consistent with an almost unchanged Young's modulus.
Nevertheless, when increasing the frequency, the branch gradually becomes orange, then yellow. This transition demonstrates the effect of the strip's lateral dimension on wave propagation and justifies solving the full 3D problem. It indicates the gradual change from a pure longitudinal polarization to a more mixed polarization. Interestingly, stretching affects the dispersion diagram when the polarization becomes predominantly transverse to the stretching direction.
In contrast, the $A_0'$ mode in a free strip is highly dependent on the applied stress, especially at low frequency where curves are purely yellow (flexion).
Similar observations can be made for other modes, in both configurations. The redder the curves, the closer together they are, so the less effect the prestress has. Conversely, the more yellow they are, the more different they are and the greater the impact of prestress.

Finally, the Dirac cone in a free strip (with symmetric modes $S_1'$ and $S_2'$) or in a fixed strip (with antisymmetric modes $A_1'$ and $A_2'$) appears orange in both configurations when the strip is undeformed. This is indeed the only case where both polarization are involved close to the $k=0$ axis. In a free strip, this Dirac cone remains orange and present. Actually, our method allows to obtain complex valued wavenumbers, and by also plotting their imaginary parts, one can notice this linear crossing is rather two straight lines avoiding each other by passing through the complex plane. By increasing the stretch ratio, cut-off frequencies almost coincide and the branches just avoid each other more (meaning that their imaginary part is slightly larger), but the Dirac cone still looks present.
In a strip with fixed edges, cut-off frequencies become sufficiently different so that the cone splits into two parts, one red which barely changes with the prestress and another one yellow which is significantly changed. In between, wavenumbers are predominantly imaginary.

\end{document}